\documentclass[pre,twocolumn,reprint,fleqn,showpacs,showpacs,showkeys]{revtex4}
\usepackage{graphicx}
\usepackage{amssymb}
\usepackage{amsmath}
\usepackage{bm}

\begin{document}

\newcommand{\bmx}{{\bm x}}
\newcommand{\bmy}{{\bm y}}
\newcommand{\bmf}{{\bm f}}
\newcommand{\bmg}{{\bm g}}
\newcommand{\bmh}{{\bm h}}
\newcommand{\bmF}{{\bm F}}
\newcommand{\bmG}{{\bm G}}
\newcommand{\calS}{{\mathcal S}}
\newcommand{\calQ}{{\mathcal Q}}
\newcommand{\bbS}{{\mathbb S}}
\newcommand{\bbQ}{{\mathbb Q}}
\newcommand{\be}{\begin{equation}}
\newcommand{\ee}{\end{equation}}

\setcounter{page}{0}
\title{Stochastic Echo Phenomena in Nonequilibrium Systems}

\author{Jae Dong \surname{Noh}}
\affiliation{Department of Physics, University of Seoul, Seoul 130-743, 
Korea}
\affiliation{School of Physics, Korea Institute for Advanced Study,
Seoul 130-722, Korea}

\date{\today}

\begin{abstract}
A thermodynamic system is driven out of equilibrium 
by a time-dependent force or a nonconservative force 
represented with a protocol $\lambda(t)$.
Dynamics of such a system is irreversible so that
the ensemble of trajectories under a time-reversed protocol 
$\lambda(-t)$ is not equivalent to that of time-reversed trajectories under 
$\lambda(t)$.
We raise a question whether one can find a suitable protocol under which the
system exhibits time-reversed motions of the original system. 
Such a phenomenon is referred to as a stochastic echo phenomenon.
We derive a condition for the
optimal protocol that leads to the stochastic echo phenomenon 
in Langevin systems. 
We find that any system driven by time-independent nonconservative
forces has a dual system exhibiting the stochastic echo
phenomenon perfectly. The stochastic echo phenomena are also demonstrated 
for harmonic oscillator systems driven by time-dependent forces.
Our study provides a novel perspective on the time-irreversibility of
nonequilibrium systems.
\end{abstract}

% 05.70.Ln : Nonequilibrium and irreversible thermodynamics
% 02.50.-r : Probability theory, stochastic processes, and statistics
% 05.40.-a : Fluctuation phenomena, random processes, noise, and Brownian motion


\maketitle

\section{Introduction }

Dynamical systems with time-reversal symmetry exhibit an echo phenomenon.
A classical mechanical system follows a time-reversed trajectory 
when the velocity of all particles is reversed. So does a quantum mechanical 
system when the Hamiltonian $H$ is switched to $-H$.  
An echo phenomenon also occurs in a driven macroscopic system.
One of the most popular examples is a drop of dye placed in a highly viscous 
fluid filled in the gap between two concentric cylinders~\cite{Taylor66}. 
When the inner cylinder is rotated, the drop spreads apparently uniformly in
the fluid. Surprisingly, the drop reappears when the rotation is reversed. 
In this system the high viscosity is the reason for the echo phenomenon. 
It prevents thermalization within an experimental time 
scale~\cite{Pine05,Corte08,During09}. 
It is an interesting question whether one can observe an echo phenomenon 
in fully stochastic systems.

An equilibrium system with Langevin dynamics obeys the detailed balance 
condition~\cite{Gardiner09} 
\be\label{DB}
p_{eq}(\bmx_i) P(\bmx_f,t_f|\bmx_i,t_i) = p_{eq}(\bmx_f)
P(\bmx_i,t_f|\bmx_f,t_i) \ ,
\ee
where $p_{eq}$ denotes the equilibrium probability density function~(PDF) and
$P(\bmx',t'|\bmx,t)$ denotes the transition probability from 
$\bmx$ at time $t$ to $\bmx'$ at time $t'$. 
The detailed balance implies that any path or trajectory $\bmx(0\leq t\leq
\tau)$ and its time-reversed one $\bmx^R(t) \equiv 
\bmx(\tau-t)$ are equally probable. 
Hence, an equilibrium system displays a stochastic echo phenomenon~(SEP),
an echo phenomenon in the probabilistic sense.

When a thermodynamic system is exerted by a time-dependent force or a
nonconservative force, it is driven into a nonequilibrium
state~\cite{Seifert05,Hatano01,Kwon11}.
We will represent a nonequilibrium driving with a protocol $\lambda(t)$,
which may depend on time or not. A nonequilibrium system does not obey the
detailed balance. So, a path $\bmx(t)$ under a 
protocol $\lambda(t)$ and its time-reversed path
$\bmx^R(t)$ under a time-reversed protocol 
$\lambda^R(t)\equiv \lambda(\tau-t)$ are not equally probable. 
In fact, the difference in the statistical weights between them is directly
related to the entropy production, which is the basis for the various 
nonequilibrium fluctuation theorems in stochastic 
thermodynamics~\cite{Jarzynski97,Kurchan98,Crooks99,Lebowitz99,Hatano01,Evans02,Seifert05,Kwon11,Noh12}. 

Due to the absence of the detailed balance, one does not expect the SEP
with a time-reversed protocol. It raises a question
whether one can control a nonequilibrium system with a suitable protocol 
to exhibit the SEP. In other words, is it possible to find an optimal
protocol under which the system evolves in time as the original system in
the time-reversed direction?
In this paper, we derive the condition for the optimal protocol for the SEP
in nonequilibrium systems following the Langevin dynamics. 
We begin with a review on stochastic thermodynamics for nonequilibrium
Langevin systems in Sec.~\ref{sec2}. 
Then, the optimal condition for the SEP will be derived in Sec.~\ref{sec3}.
The SEP will be further investigated in systems driven by nonconservative 
forces in Sec.~\ref{sec4} and by time-dependent forces in
Sec.~\ref{sec5}.
We conclude the paper with summary and discussions in Sec.~\ref{sec6}.

\section{Langevin dynamics}\label{sec2}
We consider a system 
of coordinate $\bmx = (x_1,\cdots,x_N)$ which is in thermal contact with a
heat reservoir at temperature $T$ and exerted by a force
$\bmf_{\lambda(t)}(\bmx)$. The force includes a time-independent
nonconservative force or a conservative force with a time-dependent external
parameter. Such a non-equilibrium force is represented with a protocol
$\lambda(t)$.
The Langevin equation reads~\cite{Gardiner09}
\begin{equation}\label{LangevinEq}
\dot{\bm{x}}(t) = \bmf_{\lambda(t)}(\bm{x}(t)) + \sqrt{2T}\ 
{\bm \eta}(t) \ ,
\end{equation}
where ${\bm\eta}(t)$ is the $\delta$-correlated thermal noise with
$\langle {\eta_i(t)} \rangle = 0$ and 
$\langle {\eta_i(t)} \eta_j(t') \rangle = \delta_{ij} \delta(t-t')$.
The damping coefficient and the Boltzmann constant are set to 1.

Given the initial PDF $p_i(\bmx(0))$ for $\bmx(0)$, 
the dynamics determines the PDF 
$p_f(\bmx(\tau))$ for $\bmx(\tau)$ at time $t=\tau$. 
The probability density functional for a path $\bmx(t)$ in the interval
$0\leq t \leq \tau$ is given by~\cite{Onsager53,Lau07}
\begin{equation}\label{P_lambda}
\mathcal{P}_{\lambda,p_i}[\bmx] = p_i(\bmx(0)) 
e^{-\mathcal{L}[\bmx;\bmf_\lambda]} 
\end{equation}
with the action 
\begin{equation}\label{action}
\mathcal{L}[\bmx;\bmf_\lambda] = \int_0^\tau dt \left\{
 \frac{\left[ \dot{\bmx} - \bmf_{\lambda}(\bmx) \right]^2}{4T} + 
 \frac{1}{2} \bm\nabla_\bmx\cdot{\bmf}_{\lambda}(\bmx) \right\} \ .
\end{equation}
For convenience, we adopt the Stratonovich convention for the stochastic 
integral throughout the paper~\cite{Gardiner09}. 

Suppose that the system follows a path $\bmx(t)$ under a protocol
$\lambda(t)$. The total entropy production along the path is 
related to the irreversibility as~\cite{Kurchan98,Seifert05,Kwon11}
\be\label{s_tot}
\Delta \calS_{tot}[\bmx] = \ln \left[ \frac{\mathcal{P}_{\lambda,p_i}[\bmx]}
                              {\mathcal{P}^R_{\lambda^R,p_f}[\bmx]}\right] \ ,
\ee
where $\mathcal{P}^R_{\lambda^R,p_f}[\bmx] \equiv 
\mathcal{P}_{\lambda^R,p_f}[\bmx^R]$ denotes the probability density functional 
for a time-reversed path $\bmx^R(t)\equiv \bmx(\tau-t)$ 
under a time-reversed protocol $\lambda^R(t)\equiv \lambda(\tau-t)$.
It is decomposed as 
$\Delta \calS_{tot}[\bmx] = 
\Delta \calS_{sys}[\bmx] + \Delta \calS_{res}[\bmx]$ with
\be\label{s_sys}
\Delta \calS_{sys}[\bmx] = -\ln
\left[\frac{p_f(\bmx(\tau))}{p_i(\bmx(0))}\right]
\ee
and
\be\label{s_res}
\Delta \calS_{res}[\bmx] = \mathcal{L}[\bmx^R;\lambda^R]
                           -\mathcal{L}[\bmx;\lambda] \ . 
\ee
Note that $\Delta\calS_{sys}$ is the change in the Shannon entropy of the system 
and $\Delta \calS_{res}[\bmx] = {\mathcal{Q}[\bmx]}/{T}$
is the entropy change of the heat reservoir with the heat
\be\label{heat}
\mathcal{Q}[\bmx] =  \int_0^\tau dt\ \dot\bmx(t) \cdot
\bmf_{\lambda(t)}(\bmx(t)) \ .
\ee
The quantity $-\mathcal{Q}[\bmx]$ is equal to the work done by the damping 
force and the random force. So, $\mathcal{Q}[\bmx]$ is the 
heat dissipated into the heat reservoir. 

It is noteworthy that the average value of $\Delta \calS_{tot}$ is 
the relative entropy~(or Kullback-Leibler divergence) 
of $\mathcal{P}_{\lambda,p_i}[\bmx]$ 
with respect to $\mathcal{P}^R_{\lambda^R,p_f}[\bmx]$~\cite{Cover06}: 
\be\label{s_tot_KL}
\langle \Delta \calS_{tot}\rangle_{\mathcal{P}_{\lambda,p_i}} = 
D\left({\mathcal{P}}_{\lambda,p_i} \| {\mathcal P}^R_{\lambda^R,p_f}\right) \ .
\ee
The relative entropy of a PDF $p(x)$ with respect to another PDF 
$q(x)$ is defined as 
$D(p\|q) \equiv \langle \ln \frac{p(x)}{q(x)}\rangle_{p} =
\int dx\ p(x) \ln \frac{p(x)}{q(x)}$. 
It is nonnegative for any $p$ and $q$, and 
$D(p\|q)=0$ if and only if $p(x)=q(x)$ almost everywhere.
So the entropy production vanishes only for equilibrium systems satisfying
the detailed balance. In nonequilibrium systems, the path probabilities are
not equal and the entropy production is always positive.

\section{Stochastic echo phenomenon}\label{sec3}
In order to find the optimal protocol for the SEP, 
we introduce an virtual system subject to a force 
$\bmf_{\kappa(t)}(\bmx)$ characterized by a protocol $\kappa(t)$.
The PDF for $x(0)$ of this system is taken to be
$p_f$ that corresponds to the PDF of the original system with $\bmf_\lambda$ at time $\tau$.
Then, the probability density functional for a path 
$\bmx(t)$ is given by
$\mathcal{P}_{\kappa(t),p_f}[\bmx(t)] = p_f(\bmx(0)) e^{-\mathcal{L}[\bmx;
\bmf_{\kappa}]}$.

The entropy production in Eq.~(\ref{s_tot_KL}) suggests that the
dissimilarity between the paths of the original and the virtual systems under
the time reversal can be measured by the relative entropy
\be\label{KLD}
\bbS[\kappa;\lambda]  =
D \left(\mathcal{P}_{\lambda,p_i} \|  \mathcal{P}^R_{\kappa,p_f} \right)  \
.
\ee
The optimal protocol is found by minimizing the relative entropy
with respect to $\kappa$. 
The relative entropy is useful 
since it is nonnegative and zero only for the {\em perfect} SEP with
$\mathcal{P}_{\lambda,p_i}[\bmx] = \mathcal{P}^R_{\kappa,p_f}[\bmx] =
\mathcal{P}_{\kappa,p_f}[\bmx^R]$. It reduces to the total 
entropy production for the special choice $\kappa=\lambda^R$. 
Consequently, it provides a lower bound for the total 
entropy production, 
$\langle \Delta S_{tot}\rangle  = \bbS[\lambda^R;\lambda] 
\geq \min_{\kappa} \bbS[\kappa;\lambda]\geq 0$.

Before proceeding, we remark that the optimal protocol has also been 
considered in a different context.
Suppose that a system is driven by a time-dependent external parameter
$\lambda(t)$ under the constraint that $\lambda(0) = \lambda_i$ and
$\lambda(\tau)=\lambda_f$. 
One may seek for the optimal control of $\lambda(t)$ 
that yields the minimum entropy production~\cite{Schmiedl07,Aurell11}. 
Note that this optimal protocol for the minimum entropy production is 
different from the optimal protocol for the SEP.

For a notational simplicity, we introduce
$\mu(t) \equiv \kappa^R(t) = \kappa(\tau-t)$.
Then, the relative entropy is rewritten as
\be\label{target_ftn}
\bbS[\kappa;\lambda] =  
\left\langle
\Delta\calS_{sys}[\bmx(t)]\right\rangle_{\mathcal{P}_{\lambda,p_i}}
+\frac{1}{T} \left\langle \bbQ[\bmx] \right\rangle_{
\mathcal{P}_{\lambda,p_i}} \ ,
\ee
where
\begin{eqnarray}\label{pseudo-heat}
\bbQ[\bmx] &=& -\mathcal{L}[\bmx;\bmf_\lambda] + 
               \mathcal{L}[\bmx^R;\bmf_\kappa] \nonumber \\
           &=& \int_0^\tau dt \left[ \dot{\bmx} \cdot 
               (\bmf_\lambda+\bmf_{\mu})/{2}
               - \left(\bmf_\lambda^2 - \bmf_{\mu}^2 \right)/4
               \right. \nonumber \\
           && \quad\quad\quad 
              \left. -T \bm\nabla_\bmx \cdot \left( \bmf_\lambda - 
              \bmf_{\mu} \right)/2 \right] 
\end{eqnarray}
with $\bmf_\lambda = \bmf_{\lambda(t)}(\bmx(t))$ and $\bmf_{\mu} =
\bmf_{\kappa(\tau-t)}(\bmx(t))$.
Deriving the second equality in Eq.~(\ref{pseudo-heat}), we made
a change of a variable $t \to \tau-t$ for $\mathcal{L}[\bmx^R;\bmf_\kappa]$.
The first term $\langle \Delta
\calS_{sys}\rangle_{\mathcal{P}_{\lambda,p_i}}$ 
is independent of $\mu(t)$. 
So it suffices to minimize the average of $\bbQ$. 
The functional $\bbQ$ will be called a {\em pseudo-heat} since 
it reduces to the physical heat $\mathcal{Q}$ 
for a particular choice $\mu(t)=\lambda(t)$~(see Eq.~(\ref{heat})). 

The optimal protocol $\mu_{op}$ is obtained from the condition 
$\delta \langle \bbQ\rangle/
\delta \mu(t)|_{\mu_{op}}=0$, which yields 
\be\label{optimality}
\left. \left \langle \dot\bmx \cdot 
\frac{\partial \bmf_{{\mu}}}{\partial \mu} 
+ \bmf_{{\mu}} \cdot 
\frac{\partial \bmf_{{\mu}}}{\partial \mu} 
 + T \bm\nabla_\bmx \cdot \frac{\partial
\bmf_{{\mu}}}{\partial \mu}
\right\rangle_{\mathcal{P}_{\lambda,p_i}}\right|_{\mu_{op}} = 0  \ .
\ee
To proceed further, we consider the decomposition of the form
\be\label{force_decomp}
\bmf_\lambda(\bmx) = \bmg(\bmx) + \lambda(t)\bmh(\bmx) 
\ee
with auxiliary force fields $\bmg$ and $\bmh$. The physical meaning of the
decomposition will be explained later.
In this case, the SEP is achieved with the force $\bmf_{\mu_{op}^R(t)}(\bmx) = 
\bmg(\bmx) + \mu_{op}(\tau-t) \bmh(\bmx)$ with the optimal protocol 
\be
\mu_{op}(t) = - \frac{\left\langle (\dot\bmx(t) + \bmg(\bmx(t)) + 
T \bm{\nabla}_\bmx) \cdot \bmh(\bmx(t)) \right\rangle_{{\mathcal
P}_{\lambda,p_i}}}{\langle \bmh(\bmx(t))^2 \rangle_{{\mathcal P}_{\lambda,p_i}}} \ .
\ee 
In the Stratonovich calculus, $\dot{\bmx}(t) \cdot \bmF(\bmx(t))$ should be 
interpreted as $\lim_{\delta t\to 0} \frac{1}{\delta t}
(\bmx(t+\delta t)-\bmx(t)) \cdot \frac{1}{2}(\bmF(\bmx(t+\delta
t))+\bmF(\bmx(t)))$ for any vector field $\bmF(\bmx)$~\cite{Gardiner09}.
Using the Langevin equation, one finds that
\be\label{st_rule}
\langle \dot\bmx \cdot \bmF(\bmx) 
\rangle_{\mathcal{P}_{\lambda,p_i}} = \langle (\bmf_\lambda+ T {\bm\nabla}_\bmx)
\cdot \bmF(\bmx)  \rangle_{\mathcal{P}_{\lambda,p_i}} \ .
\ee
It further simplifies the optimal condition to the form
\be\label{optimum_l}
\mu_{op} = \lambda - 2 \frac{ \langle \dot\bmx \cdot \bmh 
\rangle_{{\mathcal P}_{\lambda,p_i}}}{\langle \bmh^2\rangle_{{\mathcal
P}_{\lambda,p_i}}}
= - \lambda - 2 \frac{ \left\langle \left( \bmg + T
\bm{\nabla}_\bmx \right) \cdot \bmh \right\rangle_{\mathcal{P}_{\lambda,p_i}}}
{\langle \bmh^2 \rangle_{\mathcal{P}_{\lambda,p_i}} } \ ,
\ee 
where function arguments are omitted for simplicity.
Hereafter, the average $\langle \cdot \rangle$ is to be taken with respect 
to $\mathcal{P}_{\lambda,p_i}$, unless stated otherwise. 
Note that
$\delta^2\langle \bbQ\rangle /\delta \mu^2|_{\mu_{op}} = 
\langle \bmh^2\rangle \geq 0$. Hence, $\delta\langle \bbQ\rangle / 
\delta \mu|_{\mu_{op}}=0$ indeed provides the minimum of the pseudo-heat.

Deviation from the perfect SEP can be measured by the relative entropy 
with the optimal protocol, 
$\bbS_{op} \equiv \bbS[\kappa=\mu_{op}^R;\lambda] = 
\langle \calS_{sys} \rangle + \langle \bbQ_{op}\rangle/T$ with $\bbQ_{op} =
\bbQ|_{\mu_{op}}$. 
We define the pseudo-heat dissipation rate $\dot{\bbQ}_{op}$ 
as the integrand of Eq.~(\ref{pseudo-heat}) with $\mu=\mu_{op}$. 
Inserting in Eq.~(\ref{force_decomp}) into
Eq.~(\ref{pseudo-heat}) and using Eq.~(\ref{optimum_l}),
one can show that
\be\label{extra_heat}
\left\langle \dot{\bbQ}_{op} \right\rangle =
\left\langle \dot{\mathcal Q} \right\rangle - 
\frac{1}{4}(\lambda - \mu_{op})^2 \langle \bmh^2\rangle \ ,
\ee
where the average heat dissipation rate is given by
\be\label{dotQ}
\langle \dot{\mathcal{Q}}\rangle = \langle \dot\bmx \cdot
\bmf_\lambda\rangle = \langle (\bmf_\lambda + T \bm{\nabla}_\bmx ) \cdot
\bmf_\lambda \rangle \ .
\ee 
It is obvious that $\langle\dot{\mathcal{Q}}\rangle \geq 
\langle\dot{\bbQ}_{op}\rangle$.

In the following sections, we will apply the formalism to systems driven by
a nonconservative force and a time-dependent force, separately. 

\section{Nonconservative force case}\label{sec4}
Consider a thermodynamic system exerted by a force $\bmf(\bmx)$ which is a
sum of a conservative force $\bmf_{c}(\bmx)$ and a nonconservative 
force $\bmf_{nc}(\bmx)$. A
conservative force is given by a gradient of a potential function $V(\bmx)$ 
as $\bmf_{c}(\bmx) = -\bm{\nabla}_{\bmx} V(\bmx)$, while a nonconservative
force does not have a potential function. A nonconservative force drives a
system out of equilibrium~\cite{Hatano01,Seifert05,Kwon11}.
Along the line of Eq.~(\ref{force_decomp}), the force is decomposed as 
\be\label{f_d}
\bmf(\bmx) = \bmf_c(\bmx) + \lambda \bmf_{nc}(\bmx) 
\ee
with $\bmg = \bmf_{c}$ and $\bmh = \bmf_{nc}$.
Here, it is useful to keep $\lambda=1$ to represent the presence of a 
nonequilibrium driving.

Suppose that the system is in the nonequilibrium steady state~(NESS)
with the PDF denoted by 
$p_{ss}(\bmx) = e^{-\phi_{ss}(\bmx)}$. It is the steady state
solution of the Fokker-Planck equation~\cite{Gardiner09}
\be\label{FPeq}
\frac{\partial p}{\partial t} =
\bm\nabla_\bmx \cdot \left( -\bmf  + T \bm\nabla_\bmx \right) p  \ .
\ee
Inserting $p_{ss}(\bmx) = e^{-\phi_{ss}(\bmx)}$ into Eq.~(\ref{FPeq}), 
one finds that $\phi_{ss}(\bmx)$ should satisfy 
\be\label{ssc}
\left[ - (\bm\nabla_\bmx \phi_{ss}) + \bm{\nabla}_\bmx \right]\cdot
\left[ \bmf + T \bm\nabla_\bmx \phi_{ss} \right] = 0 \ .
\ee
In the NESS, the system entropy does not change 
while the heat is dissipated at a constant rate 
$\langle \dot{\mathcal Q}\rangle_{ss} = \langle \dot\bmx \cdot
\bmf\rangle_{ss} = \lambda \langle \dot\bmx \cdot \bmf_{nc} 
\rangle_{ss}$, where $\langle \cdot \rangle_{ss}$ 
denotes the average over the NESS~\cite{Hatano01}. 
Note that the average power of the conservative force is zero in the NESS
since  
$\langle \dot\bmx \cdot
\bmf_c \rangle_{ss} = -\frac{d}{dt}\langle V(\bmx(t))\rangle_{ss} = 0$. 

The optimal protocol in Eq.~(\ref{optimum_l}) is given by 
\be\label{op}
\mu_{op} = -\lambda - 2 \frac{ \langle (\bmf_c + T 
\bm{\nabla}_\bmx) \cdot \bmf_{nc} \rangle_{ss}}
{\langle \bmf_{nc}\rangle_{ss}^2} \ . 
\ee
It is constant in time since the averages are taken over the NESS.
From Eqs.~(\ref{extra_heat}) and (\ref{dotQ}), the average pseudo-heat 
dissipation rate is given by
\be\label{Qop_driven}
\langle \dot{\bbQ}_{op} \rangle_{ss} = \frac{ \langle \dot\bmx \cdot
\bmf_{nc}\rangle_{ss}}{\langle \bmf_{nc}^2 \rangle_{ss}} \left(
\lambda \langle \bmf_{nc}^2 \rangle_{ss} - \langle \dot\bmx \cdot
\bmf_{nc}\rangle_{ss} \right) \ .
\ee
It does not vanish in general. Therefore the relative entropy
$\bbS_{op}$ is nonzero and the SEP is imperfect.

Note that the decomposition of a given force $\bmf(\bmx)$ in 
(\ref{f_d}) is not unique. One can add and subtract a gradient of any
scalar function to the conservative force and from the nonconservative force, 
respectively~\cite{Kwon11}. 
By using this degree of freedom, one can always choose 
\be\label{simple_choice}
\bmf_c(\bmx) = - T \bm{\nabla}_\bmx \phi_{ss}(\bmx) 
\ee
using the steady-state PDF
$p_{ss}(\bmx)=e^{-\phi_{ss}(\bmx)}$.
This particular choice turns out to be extremely useful. Due to the steady
state condition in Eq.~(\ref{ssc}), $\bmf_c$ and $\bmf_{nc}$ satisfy
\be\label{ness_condition}
(\bmf_c + T \bm\nabla_\bmx) \cdot\bmf_{nc} = 0 \ .
\ee
Therefore, the optimal protocol in Eq.~(\ref{op}) is given by
\be
\mu_{op} = -\lambda \ ,
\ee
which means that the SEP is achieved by applying the nonconservative
force in the opposite direction, $\bmf = -T\bm{\nabla}_\bmx \phi_{ss} +
(\bmf+T\bm{\nabla}_\bmx\phi_{ss}) \to  -T\bm{\nabla}_\bmx \phi_{ss} -
(\bmf+T\bm{\nabla}_\bmx\phi_{ss})$.
The pseudo-heat dissipation rate can be evaluated 
from Eq.~(\ref{pseudo-heat}). Using $\mu_{op}=-\lambda$ and
Eq.~(\ref{ness_condition}), one obtains that 
$\bbQ_{op} = \dot\bmx \cdot \bmf_c = -T \frac{d}{dt} 
\phi_{ss}(\bmx(t))$. It is the total time derivative, so its
average value vanishes in the NESS.
Therefore, we conclude that a nonequilibrium system driven by a
nonconservative force can exhibit a perfect SEP. 
Precisely speaking, a system being exerted by a force 
$\bmf$ and characterized by the steady-state PDF 
$p_{ss}=e^{-\phi_{ss}}$ is equivalent to a dual system defined by 
a dual force $\bmf_{dual} = -T \bm{\nabla}_\bmx \phi_{ss} - 
(\bmf + T \bm{\nabla}_\bmx \phi_{ss}) = -\bmf(\bmx) - 
2 T \bm\nabla\phi_{ss}(\bmx)$ under the time reversal. 
Both systems share the same steady-state PDF $p_{ss}(\bmx)$.

We add a remark on the force decomposition. Suppose that the total force is
decomposed as in Eq.~(\ref{f_d}) with $\lambda=1$. It has been shown that the
special choice of $\bmf_c=-T \bm{\nabla}_\bmx \phi_{ss}$ guarantees 
Eq.~(\ref{ness_condition}).
One may ask whether the converse is also true. Suppose that the force is
decomposed as in Eq.~(\ref{f_d}) and that $\bmf_c$ and $\bmf_{nc}$ 
satisfy Eq.~(\ref{ness_condition}). Then one can show that the conservative
part is indeed equal to $\bmf_c=T\bm{\nabla}_\bmx \ln p_{ss} $ with the 
steady-state PDF $p_{ss}$ when the steady state is unique. 
The steady state condition 
along this line was discussed in Ref.~\cite{Kwon11b}.

\section{Time-dependent force case}\label{sec5}
In this case, we demonstrate the SEP with two solvable model systems.
First, consider a simple harmonic oscillator in one
dimension whose stable position is dragged according to a given 
protocol $\lambda(t)$. 
The force is given by
\be\label{dragging_force}
f_{\lambda(t)}(x) = - k (x - \lambda(t)) 
\ee
with a force constant $k$. Without loss of generality, 
we will set $\lambda(0) = 0$. This system describes a bead trapped by an
optical tweezer or an electric charge in an 
electric circuit~\cite{VanZon04,Wang02,VanZon04b,Garnier05,Imparato07}.

We assume that the particle follows the Boltzmann distribution 
$p_i(x(0))\propto e^{-kx(0)^2/(2T)}$ at $t=0$. 
The PDF for $x(t)$ at all $t$ is known exactly~(see e.g., 
Ref.~\cite{VanZon04b}). 
It follows the Gaussian distribution with the mean 
$\langle x(t) \rangle = \Lambda(t)$ and the variance 
$\langle (x(t)-\Lambda(t))^2 \rangle = T/k$ with
$\Lambda(t) = k \int_0^t ds\ \lambda(s) e^{-k(t-s)}$.

The system corresponds to the case in Eq.~(\ref{force_decomp}) with 
$g(x) = -kx$ and $h(x) = k$. Hence, the optimal protocol from 
Eq.~(\ref{optimum_l}) is given by
\be
\mu_{op}(t) = 
-\lambda(t) + 2 \langle x(t)\rangle = -\lambda(t) + 2 \Lambda(t) \ .
\ee
The PDF for $x(t)$ is given by the Gaussian 
distribution with a constant variance at all $t$. So the system entropy 
does not change. The pseudo-heat, evaluated from Eqs.~(\ref{extra_heat}) 
and (\ref{dotQ}), vanishes while $\langle \dot\calQ\rangle =
k^2(\lambda-\Lambda)^2$. 
Therefore the relative entropy $\bbS_{op}$ vanishes and the SEP is perfect. 

\begin{figure}[t]
\includegraphics*[width=\columnwidth]{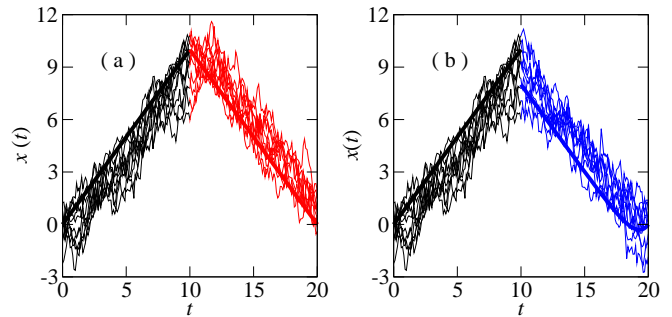}
\caption{(Color online) Sample paths of a dragged harmonic oscillator under
a protocol $\lambda$~(black, $0\leq t\leq \tau$) followed by the time-reversed
protocol $\lambda^R$~(red, $\tau\leq t \leq 2\tau$) in (a) and by
the optimal protocol  $\mu_{op}$~(blue, $\tau\leq t\leq 2\tau$) in (b). 
Thick lines represent the stable position under corresponding protocols. 
$k=T=v=1$ and $\tau=10$.}
\label{fig1}
\end{figure}

Figure~\ref{fig1} illustrates the SEP with a protocol 
$\lambda(t) = vt$. The initial position $x(0)$ is drawn from the Boltzmann 
distribution $p_i(x(0)) \propto e^{-kx(0)^2/(2T)}$. 
In the interval $0 \leq t \leq \tau$, the particle is driven under the
protocol $\lambda(t)$. Subsequently, in the interval 
$\tau \leq t \leq 2\tau$, the particle is driven under the time-reversed
protocol $\lambda(2\tau-t)$ or the optimal protocol $\mu_{op}(2\tau-t)$
where $\mu_{op}(t) = -\lambda(t) + 2 \Lambda(t) = vt - 2v(1-e^{-kt})/k$. 
Figure~\ref{fig1} shows that paths tend to lag behind the driving. Hence,
the reversibility is broken under the time reversed protocol. On the other
hand, the optimal protocol generates paths which are symmetric under the
time reversal.

As a second example, we consider a particle in one dimension subject to a force
\be\label{force_k}
f_{\lambda(t)}(x) = - \lambda(t) x 
\ee
with a time dependent force constant $\lambda(t)$~\cite{Garnier05}.
The solution of the Langevin equation is given by
$x(t) = x(0) e^{-K(t,0)} + \sqrt{2T} \int_0^t dt' e^{-K(t,t')} \eta(t')$,
where $K(t,t') \equiv \int_{t'}^t ds \lambda(s)$. 
The initial position $x(0)$ is also taken to follow the Boltzmann 
distribution $p_i(x(0)) \propto e^{-\lambda(0)x(0)^2/(2T)}$. Then,
$x(t)$, being the sum of independent Gaussian random variables, follows 
the Gaussian distribution with the mean $\langle x(t)\rangle=0$ and 
the $t$-dependent variance 
\be\label{x2}
\langle x(t)^2 \rangle = \frac{T}{\lambda(0)} e^{-2K(t,0)}
+ 2 T \int_0^t dt' e^{-2K(t,t')} \ .
\ee

This system corresponds to the case in Eq.~(\ref{force_decomp}) with 
$g(x) = 0$ and $h(x) = -x$. From Eq.~(\ref{optimum_l}), we find that the
optimal protocol is given by
\be\label{op_mu}
\mu_{op}(t) = \frac{2T }{\langle x(t)^2\rangle} - \lambda(t) \ .
\ee
In this case, the PDF for $x(t)$ changes its
shape. So the system entropy varies with time as~(see Eq.(\ref{s_sys}))
\be
\langle \Delta \mathcal{S}_{sys}\rangle = 
\langle \mathcal{S}_{sys}(t)\rangle - \langle
\mathcal{S}_{sys}(0)\rangle = \frac{1}{2} \ln \frac{ \langle x(t)^2
\rangle}{\langle x(0)^2\rangle} \ .
\ee
Differentiating it with respect to $t$ and using Eq.~(\ref{x2}), one finds 
that the system entropy increase rate is given by
\be
\langle \dot{\mathcal S}_{sys}\rangle = -\frac{\lambda(t)}{\langle x(t)^2\rangle} 
\left( \langle x(t)^2 \rangle - \frac{T}{\lambda(t)}\right) \ .
\ee
The pseudo-heat production rate calculated from Eqs.~(\ref{dotQ})
and (\ref{extra_heat}) is given by
\be
\langle \dot{\bbQ}_{op}\rangle = 
\frac{\lambda(t)T}{\langle x(t)^2\rangle} 
\left( \langle x(t)^2\rangle - \frac{T}{\lambda(t)} \right) \ .
\ee
Both $\langle \dot{\mathcal S}_{sys}\rangle$ and 
$\langle \dot{\bbQ}_{op}\rangle$ are nonzero, while 
$\langle \dot{S}_{sys}\rangle + \langle\dot{\bbQ}_{op}\rangle/T =0$.
Therefore the optimal protocol also leads to the perfect SEP.

\section{Summary and discussion}\label{sec6}
We have investigated the SEP in nonequilibrium systems driven by a
nonconservative force or a time-dependent force. 
By minimizing the relative entropy in Eq.~(\ref{target_ftn}), 
we have derived the optimal condition for the SEP
in Eq.~(\ref{optimality}). Our study shows that stochastic motions of 
nonequilibrium system can be reversed under the precise control 
using the optimal protocol. 

The harmonic oscillator system, whose stable position or force
constant is varying with time, is shown to
display the perfect SEP. It is uncertain whether the harmonic
oscillator is an exceptional case or not. Further studies are necessary to
investigate whether the perfect SEP is possible in more
complicated systems. We have shown that any nonequilibrium system 
driven by a constant nonconservative force can exhibit the perfect SEP.
Any driven system is shown to have a dual system that is 
equivalent under the time reversal. The duality may
shed some light on the general property of the NESS. The pseudo-heat plays
an important role in the characterization of time-reversibility.
The physical meaning of the pseudo-heat is an open question. 
The expression in Eq.~(\ref{Qop_driven}) suggests that the it
might be related to the violation of the fluctuation dissipation relation in
the NESS~\cite{Harada05,Prost09,Mallick11}. It should be scrutinized
further in future.

\begin{acknowledgments}
This work was supported by the Basic Science Research Program through the
NRF Grant No.~2013R1A2A2A05006776. 
We thank Prof. Hyunggyu Park and Prof. Juyeon Yi for helpful discussions.
\end{acknowledgments}


\begin{references}
\bibitem{Taylor66} G.I. Taylor, {\it Low Reynolds Number Flows} 
        (National Committee for Fluid Mechanics Films,
        Education Development Center, Newton, MA (1966)).
\bibitem{Pine05} D. J. Pine, J. P. Gollub, J. F. Brady, and A. M. Leshansky, 
        Nature {\bf 438}, 997 (2005).
\bibitem{Corte08} L. Cort\'e, P. M. Chaikin, J. P. Gollub, and D. J. Pine, 
        Nature Physics {\bf 4}, 420 (2008).
\bibitem{During09} G. D\"uring, D. Bartolo, and J. Kurchan, 
        Phys. Rev. E {\bf 79}, 030101 (2009).
\bibitem{Gardiner09} C. Gardiner, in {\em Stochastic Methods}, 4th ed. 
        (Springer, Berlin, 2009).
\bibitem{Seifert05} U. Seifert, Phys. Rev. Lett. {\bf 95}, 040602 (2005).
\bibitem{Hatano01} T. Hatano and S.-I. Sasa, 
        Phys. Rev. Lett. {\bf 86}, 3463 (2001). 
\bibitem{Kwon11} C. Kwon, J. D. Noh, and H. Park, 
        Phys. Rev. E {\bf 83}, 061145 (2011).
\bibitem{Jarzynski97} C. Jarzynski, Phys. Rev. Lett. {\bf 78}, 2690 (1997).
\bibitem{Kurchan98} J. Kurchan, J. Phys. A {\bf 31}, 3719 (1998).
\bibitem{Crooks99} G.E. Crooks, Phys. Rev. E {\bf 60}, 2721 (1999).
\bibitem{Lebowitz99} J. Lebowitz and H. Spohn, 
        J. Stat. Phys. {\bf 95}, 333 (1999).
\bibitem{Evans02} D. J. Evans and D. J. Searles, 
        Adv. Phys. {\bf 51}, 1529 (2002).
\bibitem{Noh12} J.D. Noh and J.-M. Park, Phys. Rev. Lett. {\bf 108}, (2012).
\bibitem{Onsager53} L. Onsager and S. Machlup, 
        Phys. Rev. {\bf 91}, 1505 (1953); 
        S. Machlup and L. Onsager, {\it ibid.} {\bf 91}, 1512 (1953).
\bibitem{Lau07} A. W. C. Lau and T. C. Lubensky, 
        Phys. Rev. E {\bf 76}, 011123 (2007).
\bibitem{Cover06} T. M. Cover and J. A. Thomas, 
        {\em Elements of Information Theory}, 2nd ed. 
        (Wiley, 2006).

\bibitem{Schmiedl07} T. Schmiedl and U. Seifert, 
        Phys. Rev. Lett. {\bf 98}, 108301 (2007).
\bibitem{Aurell11} E. Aurell, C. Mej\'ia-Monasterio, and 
        P. Muratore-Ginanneschi, Phys. Rev. Lett. {\bf 106}, 250601 (2011).

\bibitem{VanZon04} R. Van Zon, S. Ciliberto, and E.G.D. Cohen,
        Phys. Rev. Lett. {\bf 92}, 130601 (2004).
\bibitem{Wang02} G. Wang, E. Sevick, E. Mittag, D. Searles, and D.  Evans, 
        Phys. Rev. Lett. {\bf 89}, (2002).
\bibitem{Garnier05} N. Garnier and S. Ciliberto, 
        Phys. Rev. E {\bf 71}, 060101 (2005).
\bibitem{VanZon04b} R. Van Zon and E.G.D. Cohen, 
        Phys. Rev. E {\bf 69}, 056121 (2004).
\bibitem{Imparato07} A. Imparato, L. Peliti, G. Pesce, G. Rusciano, 
        and A. Sasso, Phys.  Rev. E {\bf 76}, 050101 (2007).
\bibitem{Kwon11b} C. Kwon and P. Ao, Phys. Rev. E 84, 061106 (2011).
\bibitem{Harada05} T. Harada and S.-I. Sasa, 
        Phys. Rev. Lett. {\bf 95}, 130602 (2005). 
\bibitem{Prost09} J. Prost, J.-F. Joanny, and J.M.R. Parrondo, 
        Phys. Rev. Lett. {\bf 103}, (2009).
\bibitem{Mallick11} K. Mallick, M. Moshe, and H. Orland, 
        J. Phys. A {\bf 44}, 095002 (2011).
\end{references}
\end{document}